\documentclass[aps,prl,twocolumn,amsmath,amssymb]{revtex4}

\begin{document}

\def\L{L}
\def\CH{{\cal H}}
\def\CO{{\cal O}}
\newcommand{\pa}{\partial}

\newcommand{\ack}[1]{[{\bf Pfft!: {#1}}]}

\preprint{hep-th/0512111}
\preprint{VPI-IPPAP-05-02}
\preprint{ILL-(TH)-05-06}

\title{Solving Pure Yang-Mills in $2+1$ Dimensions}
\author{Robert G. Leigh}
\email{rgleigh@uiuc.edu}
\affiliation{Department of Physics, University of Illinois at Urbana-Champaign\\ 
1110 West Green Street, Urbana, IL 61801, USA}
\author{Djordje Minic}
\email{dminic@vt.edu}
\author{Alexandr Yelnikov}
\email{yelnykov@vt.edu}
\affiliation{Institute for Particle Physics and Astrophysics, Department of Physics\\ 
Virginia Tech, Blacksburg, VA 24061, USA}


\begin{abstract}
We analytically compute the spectrum of the spin zero glueballs in the
planar limit of pure Yang-Mills theory in 2+1 dimensions. The new
ingredient is provided by our computation of a new non-trivial form of
the ground state wave-functional. The mass spectrum of the theory is
determined by the zeroes of Bessel functions, and the agreement with
large $N$ lattice data is excellent.
\end{abstract}

\maketitle

The understanding of the non-perturbative dynamics of Yang-Mills theory
 is one of the grand problems of theoretical physics. In this
letter we announce new analytical results pertaining to the spectrum of
the spin zero glueballs of $2+1$ dimensional Yang-Mills theory
\cite{twoplusone}. This theory is expected on many grounds to share the
essential features of its $3+1$ dimensional cousin, such as asymptotic
freedom and confinement, yet is distinguished by the existence of a
dimensionful coupling constant. Here, we determine the ground state
wave-functional in the planar limit, the knowledge of which enables us
to do the necessary computations regarding the mass gap, string tension
and the glueball spectrum. The results are in excellent agreement with
the lattice data in the planar limit \cite{teper}. Full technical
details will appear in a longer publication 
\cite{longpaper}.

Our approach, as in our previous work \cite{robme}, is based on the
remarkable work of Karabali and Nair \cite{knair}. The Karabali-Nair
approach can be summarized as follows. Consider an $SU(N)$
YM$_{2+1}$ in the Hamiltonian gauge $A_0 =0$. Write the gauge potentials
as $A_i=-it^aA_i^a$, for $i=1,2$, where $t^a$ are the Hermitian $N\times
N$ matrices in the $SU(N)$ Lie algebra $[t^a, t^b] = if^{abc} t^c$ with
the normalization $2Tr(t^at^b) =\delta^{ab}$. Define complex coordinates
$z=x_1-i x_2$ and $\bar{z} =x_1+i x_2$, and furthermore $2 A^a =
A^a_1+iA^a_2$, $2 \bar{A}^a = A^a_1-iA^a_2$.

The Karabali-Nair parameterization is
\begin{equation} \label{hl}
A= - \partial_z M M^{-1}, \quad \bar{A}= + {M^\dagger}^{-1}\partial_{\bar{z}}
M^{\dagger}
\end{equation}
where $M$ is a general element of $SL(N,{\mathbb C})$. Note that a (time
independent) gauge transformation $A \to g A g^{-1} - \partial g\
g^{-1}$, $\bar A \to g \bar A g^{-1} - \bar\partial g\ g^{-1}$, where $g
\in SU(N)$, becomes simply $M \to g M$. 
Correspondingly, a {\it local gauge invariant} variable is $H \equiv M^{\dagger} M$. 
The standard Wilson loop operator may be written
\begin{equation}
\Phi(C)=Tr P exp \{ \oint_C dz\ \partial_z H H^{-1}\} .
\end{equation}
The definition of $M$ implies a {\it holomorphic invariance} 
\begin{equation}
\begin{array}{rcl}
M(z,\bar z) &\to& M(z,\bar z)h^\dagger (\bar{z})\vspace{0.1in}\\
M^\dagger(z,\bar z) &\to& h({z})M^\dagger(z,\bar z)
\end{array}
\end{equation}
where $h(z)$ is an arbitrary unimodular complex matrix whose matrix
elements are independent of $\bar z$. This is distinct from the original
gauge transformation, since it acts as right multiplication rather than
left and is holomorphic. Under the holomorphic transformation, the gauge
invariant variable $H$ transforms homogeneously
\begin{equation}
\label{holo}
H(z,\bar z) \to h(z) H(z,\bar z) h^\dagger(\bar{z}).
\end{equation}
The theory written in terms of the gauge invariant $H$ fields will have
its own local (holomorphic) invariance. The gauge fields, and the Wilson
loop variables, know nothing about this extra invariance. We will deal
with this by requiring that the physical state wave functionals be
holomorphically invariant.

One of the most extraordinary properties of this parameterization is
that the Jacobian relating the measures on the space of connections $C$
and on the space of gauge invariant variables $H$ can be explicitly
computed \cite{knair}
\begin{equation}
d \mu [C] = \sigma d \mu [H] e^{2c_A S_{WZW}[H]}
\end{equation}
where $c_A$ is the quadratic Casimir in the adjoint representation of
$SU(N)$ ($c_A = N$), $\sigma$ is a constant determinant factor and
\begin{equation}
\begin{array}{r}
S_{WZW}(H)= - \frac{1}{2\pi}\int d^2z\ 
Tr H^{-1}\partial H H^{-1}\bar\partial H \vspace{0.1in}\\
+\frac{i}{12\pi} \int d^3x\ \epsilon^{\mu\nu\lambda} 
Tr H^{-1}\partial_\mu HH^{-1}\partial_\nu HH^{-1}\partial_\lambda H
\end{array}
\end{equation}
is the level$-c_A$ $SU(N)$ Wess-Zumino-Witten action, which is both
gauge and holomorphic invariant. Thus the inner product may be written
as an overlap integral of gauge invariant wave functionals with
non-trivial measure
\begin{equation}
\langle1|2\rangle= \int d \mu [H] e^{2c_A S_{WZW}(H)} \Psi^*_1 \Psi_2.
\end{equation}
We note that in this norm, $\Psi=1$ is normalizable.

From these expressions it is clear that a useful gauge-invariant
variable is the current
\begin{equation}
J=\frac{c_A}{\pi}\partial_z H H^{-1}
\end{equation}
which transforms as a holomorphic connection
\begin{equation}
J\mapsto hJh^{-1}+\frac{c_A}{\pi}\partial_z h h^{-1}.
\end{equation}
Note that $\bar\partial J$ transforms homogeneously, and a
holomorphic-covariant derivative is given by
\[D^{ab}=\delta^{ab}\partial_z+i\frac{\pi}{c_A}f^{abc} J^c.\]

The standard $YM_{2+1}$ Hamiltonian
\begin{equation}
{\cal H}_{YM} = \int Tr \left(g_{YM}^2 {E_i}^2 + \frac{1}{g_{YM}^2} {B}^2\right)
\end{equation}
can be also explicitly rewritten in terms of gauge invariant variables.
The collective field form \cite{collective} of this Hamiltonian (which
we will refer to as the Karabali-Nair Hamiltonian) can be easily
appreciated from its explicit form in terms of the currents as follows
\begin{widetext}
\begin{eqnarray}\label{Hamilt}
{\cal H}_{KN}[J]=T+V=
m \left(\int_x J^a(x) \frac{\delta}{\delta J^a(x)} + 
\int_{x,y}\Omega_{ab}(x,y)
\frac{\delta}{\delta J^a(x)} \frac{\delta}{\delta J^b(y)}\right) +
\frac{\pi}{m c_A} \int_x \bar{\partial} J^a \bar{\partial} J^ a
\end{eqnarray}
\end{widetext}
where 
\begin{equation}
m = \frac{g_{YM}^2 c_A }{2 \pi}, \quad \Omega_{ab}(\vec x,\vec y) = \frac{c_A}{\pi^2}
\frac{\delta_{ab}}{(x-y)^2} - \frac{i}{\pi} \frac{f_{abc}J^c(\vec x)}{(x-y)}.
\end{equation}
Interpreted as a collective field theory, one can expect to compute, {\it at large $N$}, correlators of gauge invariant operators.
Note that the magnetic field is 
\begin{equation}
B = -2\ M^{\dagger -1} \bar{\partial}(\partial H H^{-1}) M^{\dagger}
=-\frac{2\pi}{c_A}M^{\dagger -1} \bar{\partial}J M^{\dagger} .
\end{equation}
The derivation of this Hamiltonian involves carefully regulating certain
divergent expressions in a gauge invariant manner \cite{knair},
\cite{robme}. We note that the scale $m$ is essentially the 't Hooft
coupling.

The purpose of this letter is to determine masses of some of the lowest
lying glueball states. To do so, we wish to determine the form of the
vacuum wave functional and make use of the planar limit. Accordingly, we
take the following ansatz for the vacuum wave functional
\begin{equation}
\Psi_0 = \exp\left( - \frac{\pi}{2c_A m^2} \int \bar{\partial} J\ K(\L) \bar{\partial} J +\ldots\right).
\end{equation}
This form of the wavefunctional is explicitly gauge and holomorphic
invariant. The kernel $K$ is a formal Taylor expansion of $\L=(D
\bar{\partial}+\bar{\partial} D)/2m^2$, while the ellipsis contains terms higher order in $\bar\pa J$ (or $B$). This wavefunctional has the form of a ``generalized coherent state'' appropriate to large $N$ \cite{collective}, but its form is not completely dictated by large $N$ counting. The form of the ansatz, as we shall see, is sufficient to capture the mass spectrum of gauge invariant states, which we will probe using local operators. The large $N$ limit ensures that these states are non-interacting, but we are also neglecting the size of the states by using local probes. (For further details on these points, see Ref. \cite{longpaper}.)

In order to be physically sensible, $K$ should have certain properties at long and short distances. We derive these properties below.
In particular, the low momentum (large 't Hooft coupling) limit, $p^2
\ll m^2$, of the vacuum  wave functional is easily determined to be of
the form
\begin{equation}
\Psi_0 = \exp\left( - \frac{1}{2 g_{YM}^2 m} \int Tr B^2\right).
\end{equation}
(Equivalently, at low momentum, we should have $K\to 1$.) This
wavefunctional provides a probability measure $\Psi_0^* \Psi_0$
equivalent to the partition function of the Euclidean two-dimensional
Yang-Mills theory with an effective Yang-Mills coupling $g_{2D}^2 \equiv
m g_{YM}^2$. Using the results from \cite{2dym}, Karabali, Kim and Nair
deduced the area law for the expectation value of the Wilson loop
operator
\begin{equation}
\langle\Phi\rangle \sim \exp( - \sigma A)
\end{equation}
with the string tension following from the results of \cite{2dym}
\begin{equation}
\sigma = g_{YM}^4 \frac{N^2 - 1}{8\pi}.
\end{equation}
This formula agrees nicely  with extensive lattice simulations
\cite{teper}, and is consistent with the appearance of a mass gap as
well as the large $N$ 't Hooft scaling. 

Coming back to the derivation of the vacuum wave functional, we argue in Ref. \cite{longpaper} that
operators $\CO_n \equiv \int\bar{\partial} J \L^{n} \bar{\partial} J$, which would appear in a series expansion of $K(\L)$, satisfy
%
\begin{equation}
T\CO_n = (2+n) m\CO_n +\ldots
\end{equation}
In Ref. [3] (see Sec. 3 and App. A), we have presented a series of calculations supporting this important result. Further evidence is provided by lattice considerations. Given this, we can formally write
\begin{equation}
T K (\L) \longrightarrow \frac{1}{\L} \frac{d}{d\L}[\L^2 K(\L)].
\end{equation}
The full vacuum Schr\"odinger equation, combining all contributions self-consistently to quadratic order in $\bar\partial J$
\begin{equation}
\CH_{KN} \Psi_0 =E_0\Psi_0=\left[\ldots+\int tr\bar\pa J{\cal R}\bar\pa J+\ldots\right]\Psi_0,
\end{equation}
 with
suitable subtractions, then formally leads to the following differential
equation for $K$
\begin{equation}\label{Riccati}
\frac{c_Am}{\pi}{\cal R}=-K - \frac{\L}{2} \frac{d}{d\L}[K(\L)] + \L K^2 +1 =0.
\end{equation}
In this equation, the final term is the contribution of the potential
$B^2$ term of the KN Hamiltonian, while the second to last term arises
from the $\Omega$-term in the kinetic energy. Eq. (\ref{Riccati}) comes
by consistently keeping all terms quadratic in $\bar\partial J$ in the Schr\"odinger
equation.

Although this equation is non-linear, it is easily solved by
substituting $K=-y'/2y$; the resulting equation may be recast as a
Bessel equation. {\it The only normalizable solution has the correct physical asymptotics for large and
small $\L$} and is given by
\begin{equation}
K(\L) = \frac{1}{\sqrt{\L}} \frac{J_2(4 \sqrt{\L})}{J_1(4 \sqrt{\L})}
\end{equation}
where $J_n$ denotes the Bessel function of the first kind. This
remarkable formula 
encodes information on the spectrum of the theory, as
we show below. We note that this kernel has the following asymptotics
(where $\L \sim -p^2/4m^2$)
\begin{equation}
p \to 0, \quad K \to 1;\ \ \  p \to \infty, \quad K \to 2m/p
\end{equation}
consistent with confinement and asymptotic freedom, respectively.

In order to determine the spectrum, we factorize suitable correlation
functions at large distances. The operators appearing in the correlation
functions will have definite $J^{PC}$ quantum numbers, which will be
inherited by the single particle poles contributing to the correlation
function.

As a first example, we consider the $0^{++}$ states which may be probed
by the operator $Tr\bar\pa J\bar\pa J$. We have
\begin{equation}
\langle {{\rm Tr}\,(\bar{\partial} J \bar{\partial} J)}_x \ 
{{\rm Tr}\,(\bar{\partial} J \bar{\partial} J)}_y\rangle \sim \left( K^{-1}(|x-y|)\right)^2.
\end{equation}
Here, we have computed the correlation function in the planar limit
given our knowledge of the vacuum wavefunctional.

To proceed further, we note the identity
\begin{equation}
\frac{J_{\nu-1}(z)}{J_{\nu}(z)} =\frac{2\nu}{z}+ 2z \sum_{n=1}^{\infty} \frac{1}{z^2 - j^2_{\nu,n}}
\end{equation}
where $j_{\nu,n}$ are ordered zeros of the Bessel functions. For
example, the first few zeros \cite{Bessel} of $J_2(z)$ are $j_{2,1} =
5.14$, $j_{2,2} = 8.42$, $j_{2,3} = 11.62$, $j_{2,4} = 14.80$, etc.
Apart from additive constants, we then deduce
\begin{equation}
K^{-1} (k) = -\frac12 \sum_{n=1}^{\infty} \frac{M_n^2}{M_n^2+k^2}
\end{equation}
where $M_n = j_{2,n} m/2$. The Fourier transform at large $|x-y|$ is
\begin{equation}
K^{-1}(|x-y|) = -\frac{1}{4\sqrt{2\pi |x-y|}} \sum_{n = 1}^{\infty} (M_{n})^{3/2}  e^{-M_n |x-y|} .
\end{equation}

In particular the $0^{++}$ correlator mentioned above is
\begin{equation}
  \approx
\frac{1}{32 \pi |x-y|} \sum_{n,\, m =1}^{\infty} (M_n M_m)^{3/2}e^{-(M_n+M_m)|x-y|} .
\end{equation}
Note that each term here has the correct $|\vec x-\vec y|$ dependence for a single particle pole of mass $M_n+M_m$ in $2+1$ dimensions.
The $0^{++}$ glueball masses are:
\begin{equation}\label{masses}
\begin{array}{lcl}
M_{0^{++}} & = & M_1 + M_1 = 5.14 m\\
M_{0^{++*}} & = & M_1 + M_2 = 6.78 m\\
M_{0^{++**}} & = & M_2 + M_2 = 8.42 m\\
M_{0^{++**'}} & = & M_1 + M_3 = 8.38 m\\
M_{0^{++***}} & = & M_2 + M_3 = 10.02 m.\\
\end{array}
\end{equation}
Since $m$ is not a physical scale, we should re-write these results in
terms of the string tension. Given equations presented above, at large
$N$ we have $\sqrt{\sigma}\simeq \sqrt{\frac{\pi}{2}}m$. Our results are
given in Table 1.
\begin{table}
\caption{\label{Table1}$0^{++}$ glueball masses in YM$_{2+1}$. All masses 
are in units of $\sqrt{\sigma}$. 
$AdS/CFT$
computations\cite{Terning} are also given for comparison. 
The percent difference between our prediction and lattice data is given 
in the last column.}
\begin{ruledtabular}
\begin{tabular}{l|cccc}
State & Lattice, $N\to\infty$ & Sugra & Our prediction & Diff, \% \\
\hline
$0^{++}$ & $4.065 \pm 0.055$ & $4.07$(input) & $4.10$ & $0.8$ \\
$0^{++*}$ & $6.18 \pm 0.13$ & $7.02$ & $5.41$ & $12.5$\\
$0^{++**}$ & $7.99 \pm 0.22$ & $9.92$ & $6.72$ & $16$\\
$0^{++***}$ & $9.44 \pm 0.38$\footnote{Mass of $0^{++***}$ state was 
computed on the lattice for $SU(2)$ only~\cite{Meyer:2003wx}.
The number quoted here was obtained by a simple rescaling of $SU(2)$ 
result.} & $12.80$ & $7.99$ & $15$\\
\end{tabular}
\end{ruledtabular}
\end{table}
Several comments are now in order. First, note that we have been able to
predict masses of the $0^{++}$ resonances, as well as the lowest lying
member, in contrast to the original results of Karabali and Nair (which
differ significantly numerically). The supergravity results listed in
the table are a result of calculations \cite{Terning} using the AdS/CFT
correspondence \cite{adscft}; in that case, the overall normalization
was not predicted but was determined by fitting to the lattice data, for
example, to the mass of the lowest $0^{++}$ glueball. Our results for
the excited state masses differ at the 10-15\% level from the lattice
results. We note that precisely these masses are more difficult to
compute on the lattice \cite{OZ}, and thus the apparent $10-15\%$
discrepancy may be illusory.\footnote{We note also that our fourth state agrees favorably with the third lattice state, indicating that some misidentification of the lattice data may have taken place.} Finally, we note that there is some interesting approximate degeneracies in the spectrum.

Let us move on to a discussion of the $0^{--}$ glueball resonances. In
this case, our predicted masses are much closer to the lattice data,
which we believe to be more reliable in this case. We may probe these
states with the operator ${\rm Tr}\,\bar{\partial} J \bar{\partial} J
\bar{\partial} J$ .
We are thus interested in the correlation function \footnote{The $0^{--}$ probe operator used here has the correct quantum numbers but does not seem to give rise to canonically normalized poles. See Ref. \cite{longpaper} for a complete discussion.}
\begin{equation}
\langle {{\rm Tr}\,(\bar{\partial} J \bar{\partial} J \bar{\partial} J)}_x \ {{\rm Tr}\,(\bar{\partial} J \bar{\partial} J \bar{\partial} J)}_y\rangle
\sim \left( K^{-1}(|x-y|)\right)^3 .
\end{equation}
Using the results given above, we obtain glueball masses which are the
sum of three $M_n$'s.
\begin{equation}
\begin{array}{ll}
M_{0^{--}} & =  M_1 + M_1 + M_1  =  7.70 m \\
M_{0^{--*}} & =  M_1 + M_1 + M_2  =  9.34 m \\
M_{0^{--**}} & =  M_1 + M_2 + M_2  =  10.99 m. \\
\end{array}
\end{equation}
These results are compared to lattice and supergravity data in Table II.
We see that the resulting masses are within a few percent of the lattice
data, and are much better than the supergravity predictions.

\begin{table}
\caption{\label{Table2}$0^{--}$ glueball masses in YM$_{2+1}$. Columns are as in Table I.}
\begin{ruledtabular}
\begin{tabular}{l|cccc}
State & Lattice, $N\to\infty$ & Sugra & Our prediction & Diff,\%\\
\hline
$0^{--}$ & $5.91 \pm 0.25$ & $6.10$ & $6.15$ & $4$ \\
$0^{--*}$ & $7.63 \pm 0.37$ & $9.34$ & $7.46$ & $2.3$ \\
$0^{--**}$ & $8.96 \pm 0.65$ & $12.37$ & $8.77$ & $2.2$ \\
\end{tabular}
\end{ruledtabular}
\end{table}
Our results suggest that there exist hidden constituent as well as
integrable structures in $2+1$-dimensional Yang-Mills theory. Note that
the full integrability of $2+1$-dimensional pure Yang-Mills has been
suspected for some time \cite{polyakov}. In a longer paper
\cite{longpaper}, we will more carefully explain our techniques and
results and will investigate other $J^{PC}$ glueball states and the
corresponding Regge trajectories. 
It has not escaped
our attention that a similar parameterization may be used in 3+1
dimensions in a variational sense and preliminary numerical results are
encouraging.

\acknowledgements{
We would like to thank T. Takeuchi for many discussions. We would also
like to acknowledge conversations with L.~Freidel, P.~Horava,
V.~P.~Nair, J.~Polchinski and S.~G.~Rajeev. {\small RGL} was supported
in part by the U.S. DOE contract
DE-FG02-91ER40709 and {\small DM} and
{\small AY} by DOE contract DE-FG05-92ER40677.}


\begin{thebibliography}{100}

\newcommand{\wwwspires}{http://www.slac.stanford.edu/spires/find/hep/ www}

\bibitem{twoplusone}
R.~P.~Feynman,
Nucl.\ Phys.\ B {\bf 188}, 479 (1981);
I.~M.~Singer,
Commun.\ Math.\ Phys.\  {\bf 60}, 7 (1978); I.~M.~Singer,
Phys.\ Scripta {\bf 24}, 817 (1981)
P.~Orland and G.~W.~Semenoff,
Nucl.\ Phys.\ B {\bf 576}, 627 (2000)
[hep-th/9912009];
S.~G.~Rajeev,
hep-th/0401202, and references therein.

\bibitem{teper}
M.~Teper, Phys.\ Rev.\ D {\bf 59}, 014512 (1999)\ [hep-lat/9804008],
B.~Lucini and M.~Teper,
Phys.\ Rev.\ D {\bf 66}, 097502 (2002)
[hep-lat/0206027] and references therein.


\bibitem{longpaper}
R.~G.~Leigh, D.~Minic and A.~Yelnikov,
hep-th/0604060.

\bibitem{robme}
R.~G.~Leigh and D.~Minic,
hep-th/0407051.

\bibitem{knair}
D.~Karabali and V.~P.~Nair,
Nucl.\ Phys.\ B {\bf 464}, 135 (1996);
D.~Karabali and V.~P.~Nair,
Phys.\ Lett.\ B {\bf 379}, 141 (1996);
D.~Karabali, C.~j.~Kim and V.~P.~Nair,
Nucl.\ Phys.\ B {\bf 524}, 661 (1998)
[hep-th/9705087];
D.~Karabali, C.~J.~Kim and V.~P.~Nair,
Phys.\ Lett.\ B {\bf 434}, 103 (1998)
[hep-th/9804132];
D.~Karabali, C.~J.~Kim and V.~P.~Nair,
Phys.\ Rev.\ D {\bf 64}, 025011 (2001)
[hep-th/0007188] and references therein.

\bibitem{collective}
A.~Jevicki and B.~Sakita,
Nucl.\ Phys.\ B {\bf 165}, 511 (1980);
A.~Jevicki and B.~Sakita,
Nucl.\ Phys.\ B {\bf 185}, 89 (1981).
L.~G.~Yaffe,
Rev.\ Mod.\ Phys.\  {\bf 54}, 407 (1982).



\bibitem{2dym}
A. Migdal, JETP {\bf 42}, 413 (1975);
B.~E.~Rusakov,
Mod.\ Phys.\ Lett.\ A {\bf 5}, 693 (1990);
E.~Witten,
Commun.\ Math.\ Phys.\  {\bf 141}, 153 (1991); D.~S.~Fine,
Commun.\ Math.\ Phys.\  {\bf 140}, 321 (1991); M.~Blau and G.~Thompson,
Int.\ J.\ Mod.\ Phys.\ A {\bf 7}, 3781 (1992);
D.~J.~Gross,
Nucl.\ Phys.\ B {\bf 400}, 161 (1993)
D.~J.~Gross and W.~I.~Taylor,
Nucl.\ Phys.\ B {\bf 400}, 181 (1993)
[hep-th/9301068].




\bibitem{Bessel}
http://mathworld.wolfram.com/BesselFunctionZeros.html



\bibitem{Meyer:2003wx}
H.~B.~Meyer and M.~J.~Teper,
Nucl.\ Phys.\ B {\bf 668}, 111 (2003)
[hep-lat/0306019].

\bibitem{Terning}
  C.~Csaki, H.~Ooguri, Y.~Oz and J.~Terning,
  JHEP {\bf 9901}, 017 (1999)
  [hep-th/9806021];
  R.~de Mello Koch, A.~Jevicki, M.~Mihailescu and J.~P.~Nunes,
  Phys.\ Rev.\ D {\bf 58}, 105009 (1998)
  [hep-th/9806125];
  M.~Zyskin,
  Phys.\ Lett.\ B {\bf 439}, 373 (1998)
  [hep-th/9806128];
  C.~Csaki and J.~Terning,
AIP Conf.\ Proc.\  {\bf 494}, 321 (1999) [hep-th/9903142].



\bibitem{adscft}
J.~Maldacena, Adv.\ Theor.\ Math.\ Phys. {\bf 2} 231 (1998) [hep-th/9711200];
E.~Witten, Adv.\ Theor.\ Math.\ Phys. {\bf 2} 253 (1998) [hep-th/9802150]; Adv.\ Theor.\ Math.\ Phys. {\bf 2} 505 (1998) [hep-th/9803131]
S.~Gubser, I.~Klebanov and A.~Polyakov, Phys.\ Lett. {\bf B428} 105 (1998) [hep-th/9802109].


\bibitem{OZ}
  J.~Carlsson and B.~H.~J.~McKellar,
  hep-lat/0303018.

\bibitem{polyakov}
A.~M.~Polyakov,  Nucl.\ Phys.\ B {\bf 164}, 171 (1980);
{\it Gauge Fields And Strings},
Harwood, 1987.

\end{thebibliography}
\end{document}